\documentclass[12pt]{article}
\pdfoutput=1
\usepackage{cite,tabularx} 
\usepackage{amsmath}
\usepackage{amssymb}
\usepackage[top = 1. in,bottom = 0.93 in, left = 1 in, right=1 in]{geometry}

\usepackage[colorlinks=true, linkcolor=blue, bookmarks=true]{hyperref}

\makeatletter
\renewcommand\section{\@startsection {section}{1}{\z@}%
                                   {-3.5ex \@plus -1ex \@minus -.2ex}%nn
                                   {2.3ex \@plus.2ex}%
                                   {\normalfont\large\bfseries}}
\renewcommand\subsection{\@startsection{subsection}{2}{\z@}%
                                     {-3.25ex\@plus -1ex \@minus -.2ex}%
                                     {1.5ex \@plus .2ex}%
                                     {\normalfont\bfseries}}
\makeatother
\newcommand{\be}{\begin{equation}}
\newcommand{\ee}{\end{equation}}
\newcommand{\del}{\partial}
\newcommand{\sssty}{\scriptscriptstyle}

\newcommand{\dsty}{\displaystyle}
\newcommand{\s}{\star}

\newcommand{\ph}{\varphi}
\newcommand{\de}{\delta}
\renewcommand{\i}{\iota}
\newcommand{\bea}{\begin{eqnarray}}
\newcommand{\eea}{\end{eqnarray}}
\newcommand{\cL}{\mathcal{L}}
\newcommand{\cV}{\mathcal{V}}
\newcommand{\cF}{\mathcal{F}}
\newcommand{\cH}{\mathcal{H}}
\newcommand{\w}{\wedge}

\newcommand{\arXiv}[1]{\href{http://www.arXiv.org/abs/#1}{arXiv:#1}}
\newcommand{\doi}[2]{\href{http://dx.doi.org/#2}{#1}}

\begin{document}
\begin{titlepage}
\begin{flushright}
Imperial-TP-KM-2022-4
\vskip 33mm
\end{flushright}
\begin{center}
{\LARGE\bf Three approaches to chiral form interactions}    \\
\vskip 18mm
{\large Oleg Evnin$^{a,b}$ and Karapet Mkrtchyan$^c$}
\vskip 12mm
{\em $^a$ Department of Physics, Faculty of Science, Chulalongkorn University,\\
Thanon Phayathai, Pathumwan, Bangkok 10330, Thailand}
\vskip 3mm
{\em $^b$ Theoretische Natuurkunde, Vrije Universiteit Brussel (VUB) and International\\
  Solvay Institutes, Pleinlaan 2, B-1050 Brussels, Belgium}
\vskip 3mm
{\em $^c$ Theoretical Physics Group, Blackett Laboratory, Imperial College London SW7 2AZ, UK}
\vskip 7mm
{\small\noindent  {\tt oleg.evnin@gmail.com, k.mkrtchyan@imperial.ac.uk}}
\vskip 10mm
\end{center}
\vfill

\begin{center}
{\bf ABSTRACT}\vspace{3mm}
\end{center}

We briefly review and critically compare three approaches to constructing Lagrangian theories of self-interacting Abelian chiral form fields with manifest Lorentz invariance. The first approach relies on the original ideas of Pasti, Sorokin, and Tonin (PST) and has been explored since the late 1990s. The second approach was introduced by Ashoke Sen in 2015. The third approach has been developed over the last few years in the works of the present authors and other collaborators and may be called the `clone field' formalism since it features an auxiliary `clone' of the gauge field sector. We argue that this last approach shares the attractive features of the other two while avoiding their respective shortcomings. Like in Sen's approach, within the clone field formalism, arbitrary interactions can be straightforwardly included in any number of dimensions (treating interactions becomes very difficult in the PST formalism in dimensions greater than 6). Like in the PST approach, all the auxiliary fields are gauged away on-shell (while in Sen's approach, they merely decouple from the physical fields but remain dynamical).

\vfill

\end{titlepage}
%%%%%%%%%%%%%%%%%%%%%%%%%%%%%%%%%%%%%%%%%%%%%%%%%%%%%%%%%%%%%%%%%%%%%%%%%%%%%%%%%%%%%%%%%%%%%%%%%%%%%%%%%%%%%%%%%%%%%%%%%%%%%%%%%%%%%%%%%%%%%%%%%%%%%%%%%%%%%%%%%%%%%%%%%%%%%%%%%%%%%%%%%%%%%%%%%%%%%

\section{Chiral form fields}

In analogy to ordinary Abelian gauge fields, higher form gauge fields are defined by their $p$-form gauge potentials $A$ whose corresponding $(p+1)$-form field strength $F$ is given by the exterior derivative $F\equiv dA$. The free equation of motion is then $d\s F=0$, where $\s$ is the Hodge dual. (The case $p=1$ evidently recovers the ordinary Maxwell field.)

When the number of spacetime dimensions $d$ equals 2~mod~4 (with Minkowski signature), and $p=d/2-1$, it is possible to impose the selfduality relation
\be\label{freeself}
\s\! F= F,
\ee
and thus reduce the dynamical content of the theory by half.
This relation automatically implies the equation of motion $d\s F=0$, since $F$ is exact and $d$ is nilpotent. It thus describes a consistent truncation of the ordinary free theory to one-half of its degrees of freedom. Such gauge fields with selfdual field strengths are known as {\em chiral}.

While chiral gauge fields do not exist in four (Lorentzian) spacetime dimensions, they commonly appear in higher-dimensional theories explored in high-energy physics.
Thus, a chiral 4-form is encountered in ten-dimensional type IIB supergravity \cite{iib1,iib2,iib3}, and a chiral 2-form appears in the effective worldvolume description of M5-branes \cite{fivebrane1,fivebrane2} in eleven-dimensional supergravity and M-theory \cite{2ndstr1,2ndstr2,2ndstr3}.

One question that is central to our considerations is how the free selfduality relation (\ref{freeself}) could be upgraded with nonlinear terms.\footnote{While the prospect of constructing non-Abelian interactions of chiral forms \cite{nogo,nonAb} provides an important source of motivation in the context of this review, we shall in practice entirely focus on self-interactions of Abelian chiral forms. (The case of non-Abelian interactions is notoriously difficult, in particular, in view of the no-go results of \cite{nogo}).}
One could attempt to write
\be\label{FGF}
\s\! F={\cal G}( F)
\ee
with an arbitrary function $\cal G$ from $(p+1)$-forms to $(p+1)$-forms. The number of equations here is, however, the same as the number of form components, thus one expects that all degrees of freedom will be eliminated rather than the desired half. It is difficult to state an explicit condition on $\cal G$ that guarantees that half of the degrees of freedom survive. The situation becomes more manageable if one assumes that the system of algebraic equations for $F$ given by (\ref{FGF}) has been resolved to equivalently express $F-\s F$ through $F+\s F$:
\be\label{nonlchi}
F-\s F = {\cal H}(F+\s F),
\ee
where $\cal H$ is a function from $(p+1)$-forms to $(p+1)$-forms.
In this format, if only $\cal H$ satisfies, for all values of its argument, 
\be\label{sHH}
\s{\cal H}=-{\cal H}, 
\ee 
the number of independent equations drops to one-half of the number of form components. This parametrization of the general nonlinear chiral form equations was introduced in \cite{chiral} and it provides a useful perspective for the following considerations.

A well-known tough problem is constructing a Lagrangian description for chiral field equations of motion, which is challenging even for free fields. In attempts to devise such Lagrangian descriptions, difficulties arise in simultaneously maintaining locality and manifest Lorentz invariance. (Early influential constructions without manifest Lorentz invariance can be found in \cite{HT} for free chiral forms, and in \cite{PSch} for interacting ones.)

In this essay, we aim for reviewing and comparing three Lagrangian formalisms for self-interacting Abelian chiral forms with manifest Lorentz invariance: the Pasti-Sorokin-Tonin (PST) formalism originating from \cite{PST1,PST2,PST3,S5M}, the more recent formalism due to Ashoke Sen \cite{Sen1,Sen2}, and another formalism proposed in \cite{chiral} based on the prior considerations by the present authors and other collaborators in \cite{Mkrtchyan:2019opf,polynom,democ}.
This sort of analysis has in fact been invited in the conclusions of \cite{BLM}, and we are happy to provide it at this point. A brief exposition of the main features of these Lagrangian constructions will be followed by a side-by-side comparison of the three approaches. We shall argue that our formalism shares the attractive features of its predecessors while bypassing their respective shortcomings. 

Throughout the exposition, we shall be using the differential form notation, for which one can find a convenient summary in the appendices of \cite{polynom}. We mention here a few identities that are used particularly frequently in relation to the exterior derivative $d$, the wedge product $\w$, the interior product $\i_v$, and the Hodge dual $\s$. We shall be using the same letter to denote 1-forms and their corresponding vectors. For any $p$-form $A$ and a vector $v$, one has
\be\label{projrej}
\s\i_v A=(-1)^{p-1}v\w\s A, \qquad \s (A\w v)=\i_v \s A, \qquad \i_v (v\w A)+v\w\i_v A=v^2 A\,.
\ee
The last relation is known as the projection-rejection identity. The operators $d$, $v\w$, and $\i_v$ are all nilpotent, while $d$ and $\i_v$ satisfy an analog of the Leibnitz rule with respect to the wedge product. Additionally, for any two $(2k+1)$-forms $G$ and $G_1$ in $4k+2$ spacetime dimensions, one has
\be
\s\s G=G,\qquad G\w G_1=-G_1\w G, 
\ee
while for any forms $G$ and $G_1$ of equal degrees, $G\s G_1\equiv G\w\s G_1=G_1\s G$.

%%%%%%%%%%%%%%%%%%%%%%

\section{Pasti-Sorokin-Tonin approach}

In our review of the PST formalism, we follow the contemporary lucid exposition in \cite{BLM} due to
Buratti, Lechner and Melotti. In $d=4k+2$ spacetime dimensions, besides the $p=2k$-form gauge field $A$ with the field strength $F\equiv d A$, the formalism includes an auxiliary scalar $a$. The Lagrangian is
\be\label{PST-BLM}
\cL_{\sssty\mathrm{PST-BLM}}=\frac12 \,E\s B -\cV (B),\qquad E\equiv\i_v F,\qquad B\equiv\i_v\s F,\qquad v\equiv \frac{da}{\sqrt{(\del a)^2}}.
\ee
Here, $\cV$ is a scalar function of the $p$-form $B$, multiplied by the volume form. Qualitatively, one can think of $E$ as the `electric' field and $B$ as the `magnetic' field. The action becomes quadratic in the gauge field for the simple special case $\cV=B\s B/2$ that will turn out to give rise to a free chiral field. Other choices of $\cV$ result in interacting theories. It is crucial in our context to ensure that $a$ is a pure gauge degree of freedom so that there is no unwanted dynamical scalar in the theory. This will lead to a constraint on the potential function $\cV$.

Varying the Lagrangian leads to
\bea
&\dsty \de\cL_{\sssty\mathrm{PST-BLM}}=\frac12 (E\s \de B+ B\s\de E) -\de\cV (B),\qquad \s\de E=\de v\w \s F+v\w\s d\de A,\nonumber\\
& \dsty \s\de B=\de v\w  F+v\w d\de A,\qquad \de v=\frac1{\sqrt{(\del a)^2}}\i_v(v\w d\de a), \nonumber
\eea
and we write the variation of $\cV$ in the form
\be\label{varV}
\de\cV(B)= V(B)\s \de B,
\ee
which defines $V(B)$ as a function from $p$-forms to $p$-forms (it is a derivative of $\cV$ with respect to its $p$-form argument). Note that $V$ should satisfy
\be\label{ivV}
V=\i_v (v\w V),
\ee
since one has $V=\i_v (v\w V)+v\w\i_v V$ for any form $V$, while the second piece could not contribute to the variation (\ref{varV}).
Altogether, up to total derivatives, we have
\be\label{varLPSTBLM}
\de\cL_{\sssty\mathrm{PST-BLM}}=d[v\w (E-V(B))]\w \de A-\left[\frac12 (E\w E+B\w B)-V(B)\w E \right]\w v\w\de v.
\ee
(To bring the second term to the desired form one must remember (\ref{PST-BLM}), (\ref{ivV}) and the fact that any top form entirely constructed as a wedge product of factors of the form $\i_v(\ldots)$ identically vanishes.)
From this variational formula, one can see that besides the usual gauge shift of $A$ by an exact $p$-form, the theory is invariant under the gauge transformations
\be\label{PSTgauge}
\de A=da\w U +\frac{\ph}{\sqrt{(\del a)^2}}(E-V(B)),\qquad \de a=\ph,
\ee
where $U$ is an arbitrary $(p-1)$-form and $\ph$ is an arbitrary scalar, both spacetime-dependent. Validity of the $\ph$-transformation imposes a crucial condition on the derivatives of the function $\cV$ in the Lagrangian that reads
\be\label{BBVV}
B\w B=V(B)\w V(B).
\ee
It is in general difficult to analyze this relation (except for dimensions 2 and 6), which constitutes a significant hurdle for PST formalism. Because of the $\ph$-symmetry, $a$ is a pure gauge degree of freedom that can be shifted arbitrarily as long as $(\del a)^2$ remains non-vanishing.

The equations of motion are
\be\label{eomPST}
d\{v\w [E-V(B)]\}=0,\qquad d\left\{\frac{da}{(\del a)^2}\w [E-V(B)]\w[E-V(B)]\right\}=0.
\ee
The second equation is a consequence of the first and can be ignored, which is an expression of $a$ being a pure gauge degree of freedom. (To prove this, as well as the validity of the $\ph$-transformation above, one uses $dv/\sqrt{(\del a)^2}=d(1/\sqrt{(\del a)^2})\w v$. Derivations within the PST formalism generally require rather sophisticated use of exterior calculus identities. Some details of such derivations for the free case can be found in \cite{Isono}.)

The first equation in (\ref{eomPST}) can be recast as $da\w d[(E-V(B))/\sqrt{(\del a)^2}]=0$. Such equations of the form $da\w dC=0$ are generally integrated as $C=dP+da\w R$ where $P$ and $R$ are arbitrary. A derivation can be found in the appendices of \cite{polynom}. We thus get $(E-V(B))/\sqrt{(\del a)^2}=dP+da\w R$, or, acting with $da\w$,
\be
v\w(E-V(B))=da\w dP.
\ee
We can then apply the $U$-transformation of (\ref{PSTgauge}), under which $\de B=0$ and $v\w \de E=da\w dU$. One can thus always use this transformation to completely eliminate the $da\w dP$ term, leaving $v\w(E-V(B))=0$. Note that $\i_v E=0$ by construction, and $\i_v V(B)=0$ by (\ref{ivV}). Then, by the projection-rejection identity (\ref{projrej}), $v\w(E-V(B))=0$ implies
\be\label{EVB}
E=V(B),
\ee
which is the final physical equation of motion of the PST theory. This relation (nonlinear for generic choices of $\cV$) eliminates half of the field strength components, as expected for a chiral field. 

Equation (\ref{EVB}) has a curious structure. Since the equation of motion for $a$ is satisfied automatically, $a$ is an arbitrary function chosen by the end-user. Once this function has been fixed, (\ref{EVB}) is no longer Lorentz-invariant in general. It is, however, Lorentz invariant under simultaneous Lorentz transformations of $A$ and $a$. With the crucial PST invariance condition (\ref{BBVV}) respected, the $\ph$-symmetry is operational and it allows for arbitrary changes in $a$ and hence for compensating the Lorentz transformation of $a$, making the theory Lorentz-invariant under transformations of $A$ alone. From this perspective, (\ref{BBVV}) is the Lorentz invariance condition.

Since (\ref{EVB}) is Lorentz-invariant provided that (\ref{BBVV}) holds, it should be possible to recast it as (\ref{FGF}), which should presumably be of the form (\ref{nonlchi}) to leave the desired number of degrees of freedom in the theory. Developing a more explicit understanding of this transformation remains an open problem.

A significant complication for the PST formalism is that (\ref{BBVV}) is very difficult to analyze in general to identify the admissible Lagrangian functions $\cV$. The trivial solution $\cV=B\s B/2$ corresponds to free fields, where (\ref{EVB}) turns into $E=B$, which is equivalent to the free selfduality relation $F=\s F$. Interacting (Abelian) theories in this formulation are only understood fully in 2 and 6 dimensions, while (\ref{BBVV}) becomes unwieldy in higher dimensions, with only partial results available \cite{BLM}.

%%%%%%%%%%%%%%%%%%%%%%

\section{Ashoke Sen's approach}

The second approach we review was proposed in \cite{Sen1,Sen2}. While the construction drew inspiration from string field theory, the ultimate formulation is elementary and self-contained. The fundamental fields in $d=4k+2$ spacetime dimensions are a $p=2k$-form field $P$ and a selfdual $(p+1)$-form field $Q$ satisfying 
\be\label{QsQ}
Q=\s Q. 
\ee
The Lagrangian is
\be\label{LSen}
\cL_{\,\mathrm{Sen}}=\frac12\, dP\s dP -dP\w Q +\cF(Q),
\ee
where $\cF$ is a scalar function of $Q$ (multiplied by the volume form that we do not indicate explicitly). The equations of motion are
\be\label{eomSen}
d[\s dP-Q]=0,\qquad dP-\s dP= \cH(Q),
\ee
where $\cH$ is a function from selfdual $(p+1)$-forms to anti selfdual $(p+1)$-forms that is a derivative of $\cF$ with respect to its argument:
\be\label{varF}
\de\cF=\cH(Q)\w \de Q.
\ee
Note that $\cH$ satisfies
\be\label{sH-H}
\s\cH=-\cH,
\ee
since a selfdual piece in $\cH$ cannot contribute to the variation (\ref{varF}). Integrating the first equation of motion (\ref{eomSen}), we get
\be\label{sPQC}
\s dP-Q=dC,
\ee
where $C$ is arbitrary. On the other hand, applying $d$ to the second equation of motion and subtracting it from the first one yields $d[Q-\cH(Q)]=0$,
hence
\be\label{QHQ}
Q-\cH(Q)=2\,dA
\ee
for some $A$. Introducing $F\equiv dA$, and considering the Hodge dual of (\ref{QHQ}), we get, in view of (\ref{QsQ}) and (\ref{sH-H}),
\be\label{QFF}
Q=F+\s F,\qquad \cH(Q)=F-\s F,
\ee 
or
\be
F-\s F=\cH(F+\s F), 
\ee
which is the general equation of motion (\ref{nonlchi}) for interacting chiral forms; $\cH$, however, is not completely arbitrary, but arises from differentiating a scalar function as per (\ref{varF}). With (\ref{QFF}), (\ref{sPQC}) becomes
\be\label{PAB}
\s d(P-A)=dB,
\ee
where we have introduced $B\equiv C+A$,
and the second equation in (\ref{eomSen}) becomes $dP-\s dP=dA-\s dA$ or
\be\label{PAsPA}
d(P-A)=\s d(P-A).
\ee
Then from (\ref{PAB}-\ref{PAsPA}), up to an irrelevant exact term in $P$ (a gauge shift for $P$),
\be\label{PABsol}
P=A+B, \qquad dB=\s dB.
\ee
Thus, the most general solution of the equations of motion (\ref{eomSen}) is given by (\ref{PABsol}) and
\be
Q=F+\s F,\qquad F\equiv dA,\qquad F-\s F=\cH(F+\s F).
\ee
The two physical degrees of freedom are $A$, a self-interacting chiral form satisfying an equation of motion structured as (\ref{nonlchi}-\ref{sHH}) with $\cH$ being a derivative of a scalar function as per (\ref{varF}), and a decoupled free chiral form $B$ satisfying the free selfduality relation $dB=\s dB$.

Starting with a very simple Lagrangian, one thus reproduces a very large class of equations of motion for self-interacting chiral forms. One drawback is that the desired self-interacting chiral form necessarily comes with a dynamical free chiral form companion, whose associated kinetic term furthermore has a wrong sign \cite{Sen1,Sen2}.

%%%%%%%%%%%%%%%%%%%%%%

\section{The clone field approach}

Our exposition of the clone field approach follows the original presentation in \cite{chiral}. The fundamental fields in $d=4k+2$ dimensions are a $p=2k$-form gauge field $A$ with the field strength $F\equiv dA$, its auxiliary `clone' given by a $p$-form gauge field $R$ with the field strength $Q\equiv dR$, and an auxiliary scalar $a$ playing a role very similar to the $a$-field of the PST formalism. 
The Lagrangian is
\be
\mathcal{L}_{\mathrm{\,clone}}=\frac12\,H\s H+a\,F\w Q +\mathcal{F}(H+\s H),\qquad H\equiv F+aQ,
\label{lagnl}
\ee
where $\cF$ is an arbitrary scalar function of its $(p+1)$-form argument (multiplied by the volume form). The theory is invariant under the following gauge symmetries:
\begin{align}
    \de a&=0\,, \,\, \de A= dU_1\,, \,\, \de R=0\,; \label{Asymm} \\
    \de a&=0\,, \,\, \de A=0 , \,\, \de R=dU_2\,; \label{Rsymm} \\
    \de a&=0, \,\, \de A=-\,a\,da\w U\,,  \,\, \de R=da\w U\,; \label{ARgauge}\\
\de a&=\varphi\,, \,\, \de A=-\,\frac{a\,\varphi}{(\del a)^2}\,\i_{da}(Q+\s Q)\,, \,\, \de R=\frac{\varphi}{(\del a)^2}\,\i_{da}(Q+\s Q)\,,\label{PST3FQ}
\end{align}
where $U_1$, $U_2$, and $U$ are arbitrary $(p-1)$-form gauge parameters and $\ph$ is an arbitrary scalar (all the parameters are position-dependent). The first three transformations are verified straightforwardly while establishing the last one is more laborious; a derivation can be found in \cite{chiral}. Note
that $H$ is by itself invariant under (\ref{Asymm}-\ref{ARgauge}), but not under (\ref{PST3FQ}).
On the other hand, $H+\s H$ is invariant under all the symmetries (\ref{Asymm}-\ref{PST3FQ}), which automatically ensures the invariance of the interaction term $\cF(H+\s H)$. The transformation (\ref{PST3FQ}) makes it manifest that $a$ is a pure gauge degree of freedom that can be changed arbitrarily, though just like in the PST theory, $(\del a)^2$ must remain non-vanishing as it is present in the denominator of (\ref{PST3FQ}).

The equations of motion are:
\begin{align}
&d[\s H+aQ+\s X-X]=0,\label{eomchiAX}\\
&d[a\s H-aF+a\s X-aX]=0,\label{eomchiRX}\\
&Q\w\s H+F\w Q+Q\w(\s X-X)=0,\label{eomchiaX}
\end{align}
where $X$ is a derivative of $\cF$ defined by varying the interaction term so that
\be\label{varFG}
\de\cF(G)=\de G \s X(G).
\ee
Multiplying (\ref{eomchiAX}) with $a$ and subtracting it from (\ref{eomchiRX}) yields
\be\label{daHHXX}
da\w(\s H-H+\s X-X)=0.
\ee
The Hodge dual of this equation is $\i_{da}(\s H-H+\s X-X)=0$. Then, acting on this last equation with $da\w$, acting on (\ref{daHHXX}) with $\i_{da}$, and adding up the results yields, in view of (\ref{projrej}),
\be
\s\! H-H+\s X-X=0.
\label{sHX}
\ee
From this, the equation of motion for $a$ given by (\ref{eomchiaX}) is identically satisfied, as it should be, since $a$ is a pure gauge degree of freedom. Plugging (\ref{sHX}) into (\ref{eomchiAX}) yields
\be
da\w dR=0,
\ee
This is integrated in full generality as $R=dB+da\w C$ (a derivation can be found in the appendices of \cite{polynom}), which can be gauge-transformed to 
\be
R=0
\ee
 by applying (\ref{Rsymm}-\ref{ARgauge}). Thereafter, $H=F$ and, from (\ref{sHX}),
\be
\s\! F-F+\s X-X=0,
\label{FFXX}
\ee
where $X$ should now be understood as a function of $F+\s F$.

The end result is that $a$ is a pure gauge degree of freedom that can be chosen as any function with a non-vanishing $(\del a)^2$, the `clone' field $R$ has been `sacrificed,' i.e., completely gauged away, and the physical content of the theory amounts to a single chiral field $A$ satisfying the nonlinear selfduality relation (\ref{FFXX}). This selfduality relation is of the general form (\ref{nonlchi}-\ref{sHH}), where $\cH\equiv \s X-X$ is once again obtained by differentiating a scalar function as per (\ref{varFG}).

%%%%%%%%%%%%%%%%%%%%%%

\section{Comparison}

The key features of the three formalisms we have reviewed can be summarized as follows:
\begin{center}
\begin{tabular}{|m{5cm}|c|c|c|}
\hline
&PST formalism&Sen's formalism&Clone field formalism\\\hline
Interaction terms given by arbitrary functions&x&\checkmark&\checkmark\\\hline
Auxiliary fields fully gauged away&\checkmark&x&\checkmark\\\hline
Physical gauge potential among fundamental fields&\checkmark&x&\checkmark\\\hline
\end{tabular}
\end{center}
This table gives a graphical representation of our main thesis that the clone field formalism combines the strong sides of the PST and Sen's formalisms while avoiding their shortcomings.

The PST formalism is very economical in terms of its field content (only one extra auxiliary scalar), but the derivations are relatively demanding algebraically. The equations of motion are obtained in the compact form (\ref{EVB}), and if one chooses $a=t$, this form provides explicit expressions for the time derivatives of $A$ and is thus attractive, say, for evolving the initial data. The main difficulty is that the theory cannot be defined without solving first the consistency condition (\ref{BBVV}) for the functional form of the nonlinear terms. This condition appears largely intractable in dimensions higher than 6.

Sen's formalism enjoys great algebraic simplicity and produces a self-interacting chiral form satisfying very general equations (\ref{nonlchi}-\ref{sHH}), with the constraint that $\cH$ appearing on the right-hand side is obtained by differentiating an arbitrary scalar function of the form components. A drawback is that this degree of freedom is necessarily accompanied by another free chiral form, which is decoupled but remains dynamical. Another issue is that the gauge potential of the `physical' self-interacting form does not appear as an explicit variable in the theory, which may create an obstruction, say, for using this formalism as a point of departure for non-Abelian generalizations \cite{nogo,nonAb}. 

Finally, the clone field formalism generates exactly the same equations of motion as Sen's formalism for its physical degrees of freedom, while all the auxiliary fields are gauged away on-shell and do not contribute to the dynamics, as in the PST formalism. The gauge potential of the physical gauge field appears as an explicit variable in the theory.

Both Sen's formalism and the clone field formalism produce the equations of motion for their physical degrees of freedom in the simple manifestly Lorentz-covariant form (\ref{nonlchi}-\ref{sHH}). On the other hand, the physical equations of motion of the PST formalism are produced in the form (\ref{EVB}), which is not manifestly Lorentz-covariant and requires the consistency condition (\ref{BBVV}) to ensure its underlying Lorentz-covariance. We strongly suspect that, whenever (\ref{BBVV}) is satisfied, a covariant equation (\ref{nonlchi}) equivalent to (\ref{EVB}) should exist. Proving this claim does not appear straightforward, though simpler analogous connections between different forms of the equations of motion have been established for a closely related problem in four-dimensional electromagnetism in the first appendix of \cite{chiral}.

\section*{Acknowledgments}

OE is supported by Thailand NSRF via PMU-B (grants B01F650006 and B05F650021). KM is supported by the European Union's Horizon 2020 Research and Innovation Programme under the Marie Sk\l odowska-Curie Grant No. 844265 and by the STFC Consolidated Grant ST/T000791/1.

\section*{Addendum: Note on the development of the clone field formalism}

The clone field formalism first emerged in application to free (non-interacting) chiral fields in \cite{Mkrtchyan:2019opf}.
It arose while exploring consistent interactions of 2-form fields \cite{2formint}, and it was then realized that the resulting Lagrangian is related to the free PST theory by a sort of Hubbard-Stratonovich transformation.
(This straightforward relation between the PST and clone field formalisms is no longer operational when interactions are included.) The symmetries and equations of motion of the free clone theory were analyzed in more detail in \cite{polynom}, which also reported analogous constructions for democratic Lagrangians of ordinary (non-chiral) form fields, where electric and magnetic gauge potentials appear as explicit variables in the theory. Interactions were introduced into these types of theories in \cite{democ}, where a democratic description was provided for general nonlinear electrodynamics in four spacetime dimensions. Considerations of \cite{democ} suggested a natural and simple generalization to interacting $p$-forms in any number of dimensions. This generalization was explored in detail in \cite{chiral}, where the form of clone field formalism for chiral forms featured in this essay was originally developed. The follow-up work \cite{cloneiib} applied these ideas to Lagrangians of type II supergravities in ten dimensions. A supersymmetric generalization in six dimensions was derived in \cite{Kozyrev:2022dri}, while connections with the Chern-Simons approach to chiral bosons have been revealed in \cite{unif}.

%%%%%%%%%%%%%%%%%%%%%%%%%%%%%%%%%%%%%%%%%%%%%%%%%%%%%%

\end{document}